\documentclass[9pt,letterpaper]{article}
\usepackage{opex3}
\usepackage{cite}

\begin{document}
\title{Broadband large-ellipticity harmonic generation with polar molecules  }
\author{Meiyan Qin$^{1}$,Xiaosong Zhu$^{1}$, Qingbin Zhang$^{1}$, Weiyi Hong$^{1}$$^{\dag}$, and Peixiang Lu$^{1,2}$}
\address{$^1$ Wuhan National Laboratory for Optoelectronics and School of Physics, Huazhong University of
Science and Technology, Wuhan 430074, P. R. China \\
$^2$ Laboratory of Optical Information Technology, Wuhan Institute
of Technology, Wuhan 430073, P. R. China}
\email{$^\star$hongweiyi@mail.hust.edu.cn}

\begin{abstract}
We investigate the polarization properties of high harmonic
generation from polar molecules with a linearly polarized field.
It is found that elliptically polarized harmonics are observed in
a wide spectral range from the plateau to the cutoff. Further
analyses show that the nonsymmetric structure of the highest
occupied molecular orbital is the origin of ellipticity of the
harmonics. The results provide a method for generation of
large-ellipticity XUV pulses, which will benefit the application
of HHG as a tool of detection in materials and biology science.
\end{abstract}

\ocis{(190.7110) Ultrafast nonlinear optics; (190.4160)
Multi-harmonic generation; (300.6560) Spectroscope, x-ray}

\section{Introduction}
High harmonic generation (HHG) has attracted significant attention
for producing coherent attosecond pulses in the XUV regime
\cite{krausz,lan1,cao1,lan2,cao2} and detecting the molecular
structure as well as attosecond electronic dynamics
\cite{itatani,lein,lein1}. The power spectrum of high harmonic
emission has been extensively studied in previous experimental and
theoretical investigations \cite{hong1,chang,hong2}. Recently, the
polarization characteristics of the harmonic emission are also
actively studied due to the potential applications of elliptically
polarized harmonics. It is found that information about the
molecular system participating in the harmonic generation can be
imprinted in the polarization characteristics of harmonics
\cite{hijano,levesque,ramakrishna}. Therefore elliptically
polarized harmonics can provide a sensitive probe of the molecular
system. For example, Y. Mairesse {\it et al.} \cite{mairesse} both
theoretically and experimentally demonstrated that elliptically
polarized harmonics can be applied in HHG spectroscopy of oriented
N$_2$ to track multichannel dynamics during strong-field
ionization. In addition, the search for ellipticity in HHG is also
motivated by the generation of elliptically (even circularly)
polarized attosecond pulses in the xuv regime \cite{smirnova}.

Several attempts have been made to obtain large ellipticity in the
harmonics. It has been shown that the nonlinearly polarized
harmonics can be obtained using an elliptically polarized driving
laser \cite{strelkov}. In such case, the harmonic efficiency drops
significantly with increasing ellipticity of the driving laser
\cite{zhang}. The elliptically polarized harmonics are also
observed when the oriented molecule is driven by a linearly
polarized laser \cite{zhou,etches,son,le,sherratt}. Previous works
have proposed several origins of the nonzero ellipticity of the
harmonics generated from oriented molecules. For example, both
two-center \cite{etches,son} and channel \cite{smirnova}
interference, which can result in structural and dynamical minima
respectively, are responsible for the elliptically polarized
harmonics. In detail, the structural and dynamical minima in the
parallel component of the harmonics make the amplitude of the
parallel and perpendicular components comparable. Meanwhile, a
phase jump by $\sim\pi$ of the parallel component is accompanied
by the interference minimum position, leading to the phase
difference between the two components passing through $\pi/2$.
Therefore the harmonics in the vicinity of the interference minima
are elliptically polarized. Recently, Ramakrishna {\it et
al.}\cite{ramakrishna} demonstrated that the molecular potential
also plays an important role in determining the polarization
properties of the harmonics. In previous investigations, the
targets are either atoms or non-polar molecules such as H$_2$,
CO$_2$, O$_2$, and N$_2$. However, the polarization
characteristics of the harmonics generated from polar molecules
are seldom studied to the best of our knowledge. The orbital
symmetry is one of the basic properties of molecules and can
embody the difference between the polar and non-polar molecules.
Nonetheless, its influence on the elliptical polarization is also
seldom investigated.

In this paper, we investigate the polarization characteristics of
HHG from the polar molecules with a linearly polarized laser and
analyze the influence of the structure of the highest occupied
molecular orbital (HOMO). Our results show that for oriented polar
molecules strongly elliptically polarized harmonics are observed
in a spectral range from the plateau to the cutoff at the
orientation angles ranging from 30$^\circ$ to 80$^\circ$. Further
analyses reveal that the nonsymmetric structure of the HOMO is
responsible for those polarization characteristics of the
harmonics in the case of oriented polar molecules.

\section{Theoretical model}

In this paper, we use the example of CO to investigate the
\begin{figure}[htb] \centerline{
\includegraphics[width=10cm]{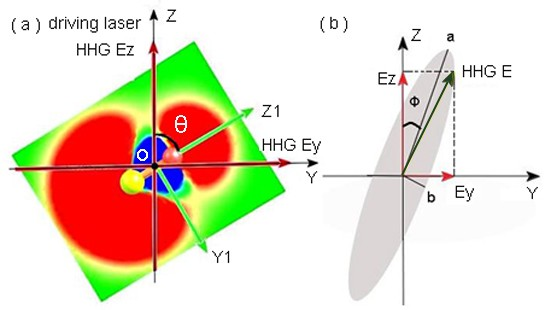}}
\caption{ (Color online) (a) An illustration of the laboratory
coordinate system. ($\mathbf{x}$,$\mathbf{y}$,$\mathbf{z}$) is the
laboratory frame,the driving laser field propagates along the
$\mathbf{x}$ axis and oscillates along the $\mathbf{z}$ axis. The
orientation angle $\theta$ is positive for clockwise rotation from
the $\mathbf{z}$ direction. (b) An illustration of HHG ellipse.
The $\mathbf{a}$, $\mathbf{b}$ axes are the major, minor axes of
the ellipse respectively. $\phi$ is the rotation angle of the
$\mathbf{a}$ axis with respect to the $\mathbf{z}$ axis.}
\end{figure}
polarization characteristics of the harmonic emission from
oriented polar molecules driven by a linearly polarized laser. Our
coordinate system, depicted in Fig. 1(a), defines the polarization
vector of the driving pulse as the space-fixed $\mathbf{z}$ axis,
with the molecular axis $\mathbf{z1}$ lying in the y-z plane. The
orientation angle between the molecular axis ($\mathbf{z1}$) and
the driving laser polarization axis ($\mathbf{z}$) is denoted as
$\theta$. Note that we only need to consider the $\mathbf{y}$,
$\mathbf{z}$ components of the harmonic emission, since the
$\mathbf{x}$ component along the propagation direction of the
driving laser cannot be phase matched \cite{zhou,le}. In Fig.
1(b), an illustration of the HHG ellipse is presented. The major,
minor axes of the ellipse are denoted as $\mathbf{a}$,
$\mathbf{b}$ respectively. $\phi$ represents the rotation angle of
the major axis ($\mathbf{a}$) of the HHG ellipse with respect to
the polarization axis ($\mathbf{z}$) of the driving laser field.
Ellipticity of HHG $\epsilon$ is defined as the ratio of the
length between the minor and major axes, i.e. $\epsilon = b/a$.

The simulations are carried out using the extended strong field
approximation (SFA) model \cite{lewen}. With its analytical and
fully quantum-mechanical formulations for harmonics, one can
explicitly and directly analyze the influence of the orbital
symmetry of the HOMO. In present work, the main focus is the
ellipticity of the harmonics, which is dominantly influenced by
the electronic dynamics. The rotational dynamics are not taken
into account by assuming a perfect orientation, which is analogous
to \cite{etches,son}. More sophisticated theory that takes both
the rotations and the molecular potential into account is
presented in recent works
\cite{ramakrishna,sherratt,ramakrishna1}. Within the SFA model,
the $\mathbf{y}$, $\mathbf{z}$ components of the laser-induced
dipole moments responsible for harmonic emission are expressed as
(in atomic units)

\begin{eqnarray}
r_{y}(t;\theta) & = & i\int^t_{-\infty}
dt'\left[\frac{\pi}{\zeta+i(t-t')/2}\right]^{3/2}E(t')g(t';\theta)d_z\left[p_{st}(t',t)+A(t');\theta\right]\nonumber\\
 & &\times exp\left[-iS_{st}(t',t)\right ]g^{*}(t;\theta)d_y^{*}\left[p_{st}(t',t)+A(t);\theta\right]+c.c..
\end{eqnarray}
\begin{eqnarray}
r_{z}(t;\theta) & = & i\int^t_{-\infty}
dt'\left[\frac{\pi}{\zeta+i(t-t')/2}\right]^{3/2}E(t')g(t';\theta)d_z\left[p_{st}(t',t)+A(t');\theta\right]\nonumber\\
 & &\times exp\left[-iS_{st}(t',t)\right ]g^{*}(t;\theta)d_z^{*}\left[p_{st}(t',t)+A(t);\theta\right]+c.c..
\end{eqnarray}
In these equations, $E(t)$ refers to the electric field of the
driving laser pulse, $A(t)$ is its associated vector potential,
$\zeta$ is a positive constant, $t'$ and $t$ correspond to the
moments of electronic ionization and recombination respectively,
$p_{st}$ and $S_{st}$ are the stationary momentum and the
quasi-classical action, which are given by
\begin{equation}
p_{st}(t',t)=\frac{1}{t-t'}\int^t_{t'}A(t'')dt'',
\end{equation}
\begin{equation}
S_{st}(t',t)=(t-t')I_{p}-\frac{1}{2}p^{2}_{st}(t',t)(t-t')+
\frac{1}{2}\int^t_{t'}A^{2}(t'')dt'',
\end{equation}
where $I_{p}$ is the ionization energy of the target molecule.
$g(t;\theta)$ represents the ground state amplitude at the time
$t$, and can be expressed as \cite{lan,hong}:
\begin{equation}
g(t;\theta) = exp \left [-\int^t_{-\infty} w(t';\theta)dt'\right].
\end{equation}
$w(t';\theta)$ is the ionization rate, which is calculated by
Molecular Ammosov-Delone-Krainov (MO-ADK) model for aligned
molecules \cite{madk}. $\vec{d}(p)$ is the field-free dipole
matrix element for transition from the ground state to the
continuum state, where p stands for the momentum of the electron.
$d_y, d_z$ correspond to its $\mathbf{y}$, $\mathbf{z}$
components, respectively. Within the single active electron
approximation (SAE), this transition dipole moment is given by
\cite{levesque}
\begin{equation} \vec{d}(p; \theta)=
\langle\psi_0(x,y,z;\theta)|\vec{r}|\psi_p\rangle.
\end {equation}
Here $\psi_0(x,y,z;\theta)$ represents the ground state of the
target molecule, i.e. the highest occupied molecular orbital
(HOMO). The HOMO is obtained using the Gaussian 03 ab initio code
\cite{gauss} with the 3-21G basis set. $\psi_p = \exp{(ipz)}$
refers to the electronic continuum state with a momentum p.
 The harmonic
spectrum is then obtained by Fourier transforming the
time-dependent dipole acceleration $\vec{a}(t;\theta)$. The
equations for the two components are:
\begin{equation}
a_y(q;\theta)=\frac{1}{T}\int^T_0a_y(t;\theta)exp(-iq\omega t)dt,
\end{equation}
\begin{equation}
a_z(q;\theta)=\frac{1}{T}\int^T_0a_z(t;\theta)exp(-iq\omega t)dt,
\end{equation}
where $\vec{a}(t;\theta)=\ddot{\vec{r}}(t;\theta)$, $T$ and
$\omega$ are the duration and frequency of the driving pulse, $q$
corresponds to the harmonic order. The spectral intensity ( $I$ )
and phase ($\varphi$) of harmonic components are given by $I_j(q)
= |a_j(q)|^2$ , $\varphi_j(q) = \arg[a_j(q)]$ , $j = y, z$. The
ellipticity $\epsilon$ is determined by the amplitude ratio and
the phase difference of the two components \cite{son}:
\begin{equation}
\epsilon =
\sqrt{\frac{1+r^2-\sqrt{1+2r^{2}\cos2\delta+r^4}}{1+r^2+\sqrt{1+2r^2\cos2\delta+r^4}}}
\end{equation}
where $r=\sqrt{I_y / I_z}$ and $\delta = \varphi_y - \varphi_z$.
The range of the ellipticity is $0\leq\epsilon\leq1$. The linear,
elliptical, and circular polarization correspond to $\epsilon=0$ ,
$0<\epsilon<1$ , and $\epsilon=1$ respectively. If $\delta$ is 0
or $\pi$, the HHG will be linearly polarized. For other values of
$\delta$ between 0 and $\pi$, the HHG will be elliptically
polarized. In order to attain high ellipticity, the phase
difference must be around $\pi/2$.

\section{Result and discussion}

\begin{figure}[htb]
\centerline{
\includegraphics[width=10cm]{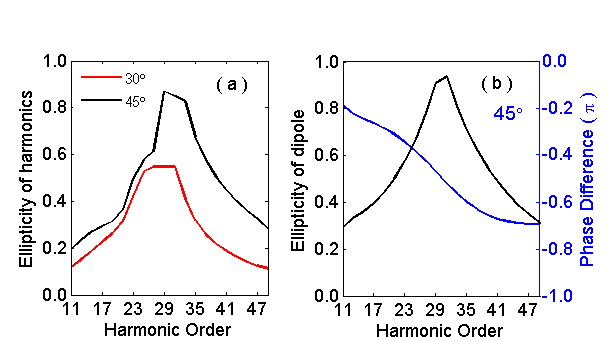}}
\caption{(Color online) (a) The ellipticity of harmonics generated
from CO for the orientation angle at $30^\circ$ (the red line) and
$45^\circ$ (the black line). The horizontal axis represents the
harmonic order. (b) The ellipticity (the black line) and phase
difference (the blue line) of the recombination dipole moment for
the orientation angle at $45^\circ$. The phase difference is
shifted into $[-\pi,0]$. The electronic momentum is transformed
into the harmonic order.}
\end{figure}
\begin{figure}[htb]
\centerline{
\includegraphics[width=10cm]{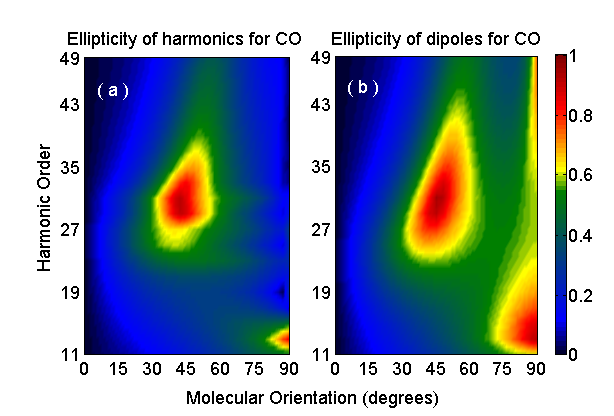}}
\caption{(Color online) (a) The ellipticity of harmonics as a
function of the orientation angle and the harmonic order. (b)The
ellipticity of the recombination dipole moments versus the
orientation angle and the electronic momentum which is transformed
into the harmonic order using the dispersion relation $q
\hbar\omega=E_p+I_p$. E$_p$ is the kinetic energy of the free
electron with the momentum p, Ip is the ionization energy of the
target. For the two panels, the target is the polar CO molecule,
the orientation angle and the harmonic order correspond to the
horizontal and vertical axes respectively.}
\end{figure}
We study the ellipticity of HHG from oriented polar CO molecules
driven by a linearly polarized laser.A 30-fs 800-nm driving pulse
with an intensity of $2\times10^{14}$ $W/cm^2$ is used in our
simulation. The electric field of the laser pulse is expressed as
\begin{equation}
\vec{E}(t) = E_0 \cos(\omega t)sin^2(\frac{\pi t}{T})\vec{z},
\end{equation}
where $E_0$ is the amplitude of the driving field. Figure.
2(a)presents the ellipticity of the harmonic emission at two
orientation angles $30^\circ$ (red line) and $45^\circ$ (black
line). One can see that for the two orientation angles, strongly
elliptically polarized harmonics are observed in a wide spectral
\begin{figure}[htb]
\centerline{
\includegraphics[width=10cm]{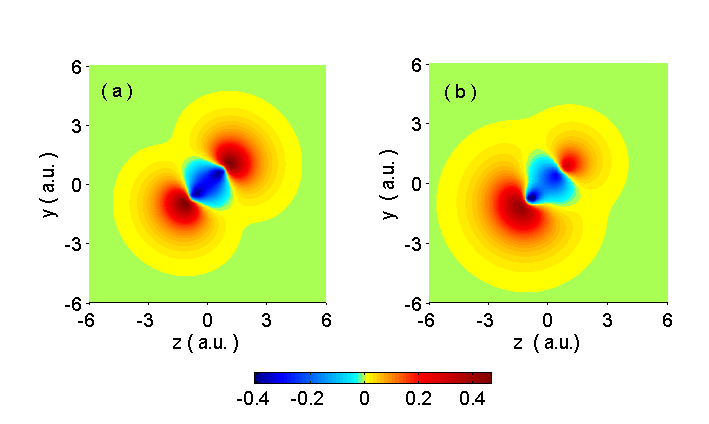}}
\caption{(Color online) A view of the section parallel to the yz
plane for N$_2$ in the left panel and CO in the right panel. The
two target molecules are oriented at $45^\circ$.}
\end{figure}
range covering the plateau to the cutoff. In Fig. 2(b), the phase
difference (blue line) and the ellipticity (black line) of the
recombination dipole moment \cite{levesque,zhao} for CO oriented
at $45^\circ$ are also presented. Here the phase difference is
shifted into [$-\pi$, 0]. Interestingly, the relative phase
between the $y$, $z$ components of the dipole moment is close to
$-\pi/2$ for all the electronic momenta. As a result, the
ellipticity of the dipole moment is quite large, as shown by the
black curve. Furthermore, it is also shown that the ellipticity
curve of the dipole moment in panel (b) is similar to that of the
harmonics at the same orientation angle (black line) in panel (a).
The electronic momentum is transformed into the harmonic order by
the dispersion relation $q \hbar\omega=E_p+I_p$, wherein E$_p$ is
the kinetic energy of the free electron with the momentum p. This
phenomenon will be discussed latter.

\begin{figure}[htb] \centerline{
\includegraphics[width=10cm]{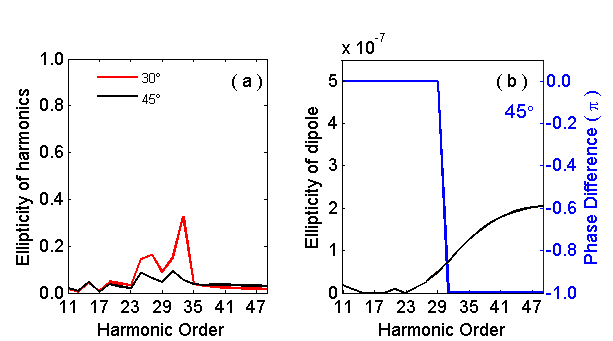}}
\caption{(Color online) (a) The ellipticity of harmonics generated
from the oriented N$_2$ at $30^\circ$ (the red line) and
$45^\circ$ (the black line). The horizontal axis is the harmonic
order. (b) The ellipticity (the black line) and phase difference
(the blue line) of the recombination dipole moment for N$_2$
oriented at $45^\circ$. The phase difference is shifted into
$[-\pi,0]$. The electronic momentum is transformed into the
harmonic order.}
\end{figure}
The ellipticity of the harmonics for CO oriented at other angles
from $0^\circ$ to $90^\circ$ is also studied. The map of the
harmonic ellipticity is shown in Fig. 3(a) versus the harmonic
order and the orientation angle. As shown in Fig. 3(a), strongly
elliptically polarized harmonics are also obtained at other
orientation angles ranging from $30^\circ$ to $80^\circ$. In Fig.
3(b), the ellipticity of the dipole moment as a function of the
orientation angle and the electronic momentum is presented as
well. Here the electronic momentum is transformed into the
harmonic order using the same dispersion relation. One can see
that the phenomenon observed in Fig. 2 is not specific to the case
when the molecule is oriented at angle $45^\circ$. As shown in
Fig. 3, the ellipticity map of the harmonics in panel (a) also
exhibits a quite similar shape to that of the dipole moment in
panel (b). To explain this phenomenon, a theoretical analysis is
performed based on the three-step model \cite{lewen,corkum} for
high harmonic generation from oriented molecules. In this model,
an electron first tunnels from the ground state, then oscillates
in the field, finally recombines with the parent ion. The
recombination step leads to the emission of a high-energy photon
with two polarization components, parallel ($z$) and perpendicular
($y$) to the laser polarization. Since the first two steps are
identical for the y, z components of the harmonics, the phase
difference and amplitude ratio between the two components of the
harmonics is mainly contributed by those of the recombination
dipole moment $\vec{d}^*(p)$. Hence the ellipticity curves between
harmonics and the recombination dipole moment exhibit a similar
shape at the same orientation angle, as shown in Fig. 2 and Fig.
3. In other words, the recombination dipole moment $\vec{d}^*(p)$
is responsible for the large ellipticity of the harmonics over a
wide spectral range for CO.

From the formula $\vec{d}^*(p; \theta)=
\langle\exp{(ipz)}|\vec{r}|\psi_0(r;\theta)\rangle$ for the
recombination dipole moment, one can see that the phase difference
$\arg(d^*_y)-\arg(d^*_z)$ and the amplitude ratio $|d^*_y/d^*_z|$
of the recombination dipole moment are directly related to the
spatial structure of the HOMO. In the following, we analyze the
influence of the HOMO structure on the phase difference of
$\vec{d}^*(p)$ in detail to grasp a deeper insight into the large
ellipticity of the harmonics generated from the polar CO
molecules. For comparison, we also discuss the case of non-polar
N$_2$ molecule, which has a symmetric HOMO with a similar geometry
($\sigma_g$) to that of CO. Sectional views of the HOMO for N$_2$
and CO at the orientation angle $45^\circ$ are presented in Fig.
4. As shown in this figure, the section of the non-polar N$_2$
molecule in panel (a) exhibits a centro-symmetric structure. Then
the wavefunction satisfies $\psi_0(x,y,z;\theta) =
\psi_0(x,-y,-z;\theta)$ (for arbitrary value of x). However, for
the nonsymmetric HOMO of CO this condition is not satisfied, as
shown in panel (b). With Eq.(6), one can obtain $Re[d^*_j]=0$ and
$Im[d^*_j]\neq0$ for N$_2$, $Re[d^*_j]\neq0$ and $Im[d^*_j]\neq0$
($j = y, z$) for CO, respectively. As a result the phase
difference of $\vec{d}^*(p)$ for N$_2$ is either $0$ or $\pm\pi$,
as shown by the blue curve in Fig. 5(b). Whereas for CO, they can
take other values including $\pm\pi/2$ for all the electronic
momenta, as demonstrated in Fig. 2(b). The phase difference of the
harmonics is dominated by that of the recombination dipole moment,
as illustrated above. Therefore large ellipticity of the harmonics
is obtained in a wide spectral spectrum, as presented in Fig. 2(a)
by the black curve. Finally, it can be concluded from the above
discussion that the nonsymmetric structure of the HOMO is the
major origin of the ellipticity observed in the case of the
oriented polar CO molecules.

Figure 5(a) shows the ellipticity of the harmonics generated from
N$_2$ at the same orientation angles with Fig. 2(a). In contrast
to CO, the nonzero ellipticity of the harmonics are only obtained
in the vicinity of the structure minima in the HHG spectra, where
the positions of the minima coincide with those in \cite{zhao}.
This nonzero ellipticity of the harmonics is well explained by the
two-center interference effects, which is analogous to
\cite{smirnova,etches,son} for non-polar molecules. Comparison
between the cases of CO and N$_2$ also shows that the ellipticity
of the harmonics of CO is larger than that of N$_2$ for most of
the orientation angles, as demonstrated in Fig. 2 and Fig. 5 for
the orientation angles at $30^\circ$ and $45^\circ$. Therefore our
results suggest a potential application of HHG from polar CO
molecules in generating the broadband large-ellipticity harmonics,
which will benefit the application of HHG as a tool of detection
in biology and materials science in the future.

\section{Conclusion}
We use the example of CO to investigate the polarization
characteristics of HHG from the oriented polar molecules driven by
a linearly polarized laser. It is found that the harmonics over a
wide spectral range are strongly elliptically polarized for the
orientation angles ranging from $30^\circ$ to $80^\circ$. The
spectral range of large-ellipticity harmonics is covered from the
plateau to the cutoff, while for non-polar molecules the harmonics
only in the vicinity of the interference minimum are elliptically
polarized. Further analyses show that the nonsymmetric structure
of the HOMO is the major origin of the ellipticity of the
harmonics observed in oriented CO molecules. Since the
nonsymmetric structure is a general characteristic of the HOMO for
polar molecules, the conclusions made from the results of CO can
be extended to other polar molecules. Thus our investigations
provide a method for generation of large-ellipticity XUV pulses,
which will benefit the application of HHG as a tool of detection
in materials and biology science.

\section*{Acknowledgment} This work was supported by the National Natural Science Foundation of China under Grants
No. 60925021, 10904045, 10734080 and the National Basic Research
Program of China under Grant No. 2011CB808103. This work was
partially supported by the State Key Laboratory of Precision
Spectroscopy of East China Normal University.

\end{document}